\documentclass[fleqn,usenatbib]{mnras}

\usepackage{newtxtext,newtxmath}

\usepackage[T1]{fontenc}

\DeclareRobustCommand{\VAN}[3]{#2}
\let\VANthebibliography\thebibliography
\def\thebibliography{\DeclareRobustCommand{\VAN}[3]{##3}\VANthebibliography}

\usepackage{graphicx}	
\usepackage{amsmath}	

\title[]{Scaling and Universality in the Temporal Occurrence of Repeating FRBs}

\author[Y. Du et al.]{
Yan-Qi Du,$^{1,2}$
Ping Wang,$^{1}$\thanks{E-mail: pwang@ihep.ac.cn}
Li-Ming Song$^{1}$
and Shao-Lin Xiong$^{1}$
\\
$^{1}$Key Laboratory of Particle Astrophysics, Institute of High Energy Physics, Chinese Academy of Science, 100049 Beijing, China\\
$^{2}$Information Science and Technology, Southwest Jiaotong University, 610031 Chengdu, China
}

\date{Accepted XXX. Received YYY; in original form ZZZ}

\pubyear{2023}

\begin{document}
\label{firstpage}
\pagerange{\pageref{firstpage}--\pageref{lastpage}}
\maketitle

\begin{abstract}
Fast Radio Bursts (FRBs) are energetic phenomena that have significant implications for understanding fundamental physics and the universe. Recent observations of FRB 121102, FRB 20220912A, and FRB 20201124A by the Five-hundred-meter Aperture Spherical Telescope (FAST) showed high burst rates and distinctive energy distribution and temporal properties. In this study, we examine these observations to investigate the scale invariance of the waiting times between bursts for intervals longer than approximately 1 second. Our analysis revealed a unified scaling law for these longer intervals, which is similar to the behavior of solar flares. This discovery inspires us to suggest a dual analogy of the FRB scenario across the entire time intervals: with earthquake dynamics at subsecond scales and with solar flare dynamics beyond the one-second threshold. This threshold potentially aligns with the dynamic time scale of neutron star crusts, offering insight of the occurrence of FRBs into the internal processes of neutron stars. 
\end{abstract}

\begin{keywords}
radio continuum: transients, fast radio bursts
\end{keywords}



\section{Introduction}

Fast Radio Bursts (FRBs) are enigmatic phenomena, ranking among the brightest and most energetic outbursts in radio wavelength in the universe. These intense millisecond-duration bursts have attracted a great deal of attention due to their potential to probe fundamental physics and the evolution of the universe \citep{Michilli, Katz, Petroff, Cordes}. With their ability to inform our understanding of cosmological parameters, track cosmic magnetic fields, test dark matter theories, and explore gravitational waves, FRBs have become a center of attention in recent astrophysical research \citep{keane_future_2018}.

Repeating FRBs, in particular, exhibit intriguing non-Poisson clustering in their burst time distribution, suggesting that the burst rate varies over time \citep{wang_sgr-like_2017, Oppermann}. This non-Poisson behavior, potentially linked to the local environment or intrinsic characteristics of the FRB source, remains a subject of intense study. The case of FRB 121102 \citep{FAST1}, the first accurately located repeating FRB source, is especially notable. Its burst energy distribution follows a power law, and large-scale statistical analyses have found no periodic or quasiperiodic signals, challenging models dependent on a single rotating compact object. Similarly, FRB 20201124A \citep{FAST2, zhouFASTObservationsExtremely2022} exhibits high activity without clear periodicity, hinting at the need for an emission mechanism with remarkable efficiency.

The phenomenon of FRBs, while distant and mysterious, shares similarities with more familiar terrestrial and solar events such as earthquakes and solar flares \citep{cheng_earthquake, boffettaPowerLawsSolar1999b, Cruces}. These phenomena all involve abrupt energy releases, although on vastly different scales and environments. Earthquakes result from tectonic stress causing faults in Earth's crust to rupture, whereas solar flares emerge from the reconnection of magnetic field lines that store energy in the corona and trigger intense bursts on the Sun's surface. Despite their different energy generation mechanisms, these events exhibit complex interactions among various physical processes and show power law energy distributions.

Recent studies have postulated that magnetic energy in magnetar magnetospheres could be a key to understanding FRBs \citep{Katz1}. The first substantial link between FRBs and magnetars was established through the simultaneous detection of gamma-ray and x-ray emissions with an FRB, suggesting that disruptions in a magnetar's magnetic field lines, possibly due to starquakes, might trigger radio emissions \citep{liHXMTIdentificationNonthermal2021, andersenBrightMilliseconddurationRadio2020, bochenekFastRadioBurst2020, zhang_physical_2020}. Interestingly, this similarity is how earthquakes, despite their different origins, share similarities with starquakes \citep{cheng_earthquake}. Therefore, insights from earthquake studies, which have revealed a unique universal distribution that governs recurrence times, may offer valuable perspectives on FRBs. Similarly, the temporal occurrence of solar flares, described within the framework of SOC, presents another comparative avenue.

Studies have demonstrated bursts clustering over time and the distribution of power law energy for FRB 121102, FRB 20201124A and FRB 20220912A \citep{Zhang, Gourdji, FAST1, FAST2, Cruces, wang_sgr-like_2017, Oppermann, zhouFASTObservationsExtremely2022, zhangFASTObservationsFRB2023}, suggesting that its occurrence is governed by a complex stochastic process. Furthermore, recent studies have revealed a bimodal distribution in waiting times between bursts. In particular, the presence of a shorter peak within this distribution, ranging from a few milliseconds to several tens of milliseconds, was consistently observed across all four burst source datasets. This may be attributed to the particular criteria adopted to define individual bursts \citep{FAST1, FAST2, zhangFASTObservationsFRB2023, zhouFASTObservationsExtremely2022}. Further investigations, such as those conducted by \cite{totaniFastRadioBursts2023}, have demonstrated the short-time correlation within the shorter peak, while the longer peak can mostly be described by an uncorrelated Poisson process. The study also revealed that repeating FRBs exhibit characteristics similar to earthquakes on a timescale shorter than 1 s, suggesting seismic-like activity in neutron star crusts as the underlying mechanism.

Our study focuses on the temporal clustering of repeating FRBs, using high-quality data sets from FRB 121102 \citep{FAST1}, comprising 1652 independent bursts over $59.5$ hours spanning $47$ days, and FRB 20201124A \citep{FAST2}, with $1863$ bursts over $82$ hours across $54$ days. Another is related to FRB 20220912A, covering $1,076$ bursts over $8.67$ hours, spanning about $55$ days \citep{zhangFASTObservationsFRB2023}. The last data set records another intensely active phase of FRB 20201124A, with $881$ bursts recorded approximately within $4$ hours over 4 days \citep{zhouFASTObservationsExtremely2022}. All data sets, recorded by FAST, offer unprecedented resolution and lower flux thresholds. To distinguish between the two, this paper will refer to the data sets associated with FRB 20201124A sequentially as FRB 20201124A (A) and FRB 20201124A (B) in the subsequent sections. Our analysis investigates the waiting times and their scale invariance across different fluence thresholds. We present compelling evidence of a novel scaling and universality in these waiting times, which was not reported before. Our results demonstrate a unified scaling law, valid over three orders of magnitude, and reveal a surprising similarity in functional form to solar flares rather than earthquakes, which is complementary to \cite{totaniFastRadioBursts2023}, where the authors argue that the repeating FRBs temporal behavior on a short time scale is more like earthquakes than solar flares. This finding hints at a dynamical complexity in the occurrence of FRB.

Given the discontinuous nature of FAST's observations of FRB 121102, FRB 20201124A, and FRB 20220912A, we define waiting time as the interval between successive bursts within the same observation period, thereby avoiding long observational gaps. We describe the temporal clustering of FRBs through the recurrence interval distribution $P(\Delta t)$, where $\Delta t = t_{i+1} - t_i$ represents the interval between the $i$th and $(i + 1)$th bursts. For our statistical analysis, we adopt logarithmic bins to ensure appropriate sizing for each timescale, although the probability density function (PDF) calculated renders the specific number and width of bins less critical to our analysis.

\section{Results}

\begin{figure}
 \centering
 \includegraphics[width=\columnwidth]{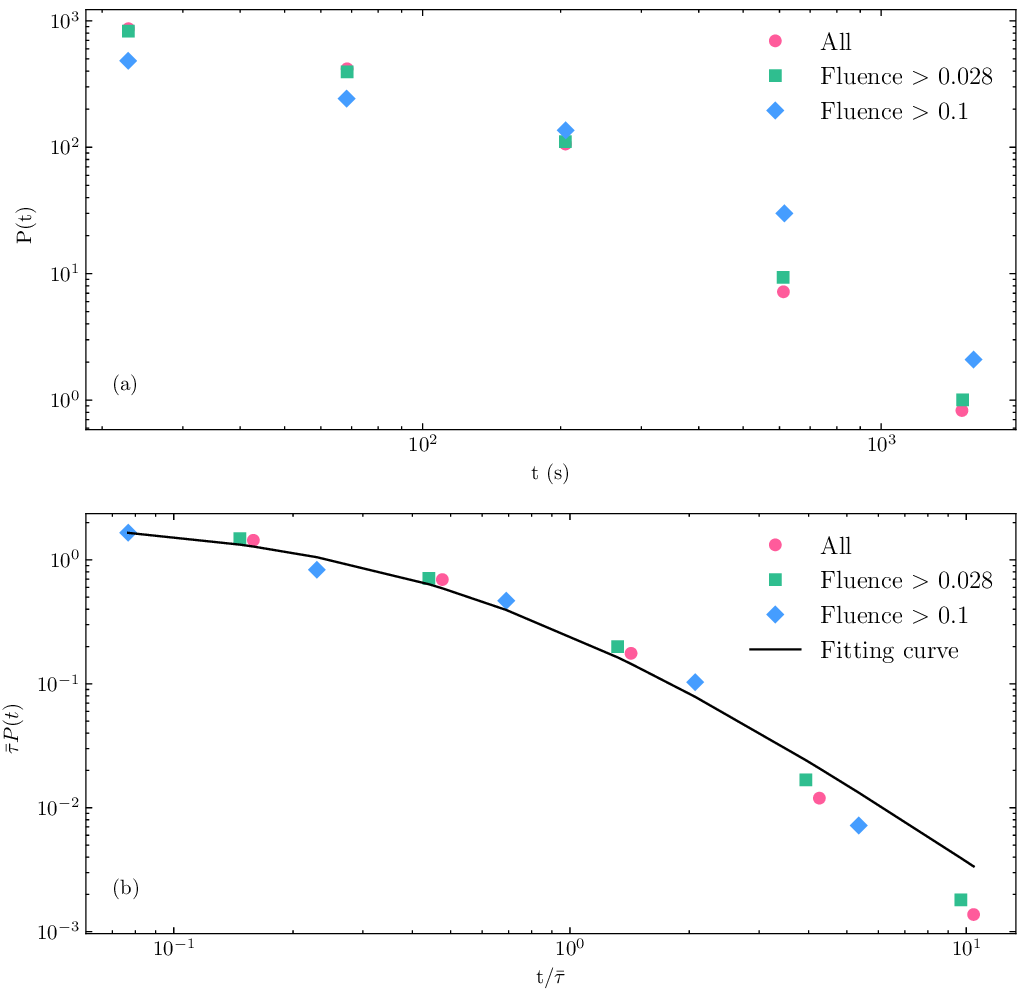}
 \caption{Waiting time distributions without and with rescaling of FRB 121102. (a) Waiting time distributions $P(t)$ for different fluence thresholds (Unit: Jy ms). One of the thresholds of fluence 0.281 is approximately near to energy $3.0 \times 10^{38}$ (erg). (b) Waiting time distributions after rescaling by the average waiting time of bursts in the case of different fluence thresholds, respectively.}
 \label{fig:frb121102}
\end{figure}

The characteristics of the waiting time distributions $P(t)$ of FRB 121102 are shown in Figure \ref{fig:frb121102}(a). For each threshold, the range of PDF bins was set between the minimum and maximum values of $\Delta t$ in equal logarithmically spaced bins. Here, $P(t)$ exhibits a power law tail with different slopes for different intensity thresholds. With an increasing threshold, $P(t)$ evolves smoothly, becoming flatter up to longer times. All of them exhibit similar behavior, with a slow decrease at the beginning of their temporal range and ending with a much sharper decay.

The scaling law for the waiting time distribution was first proposed by Bak et al. \citep{Bak_wt} and later by Corral \citep{corralLongTermClusteringScaling2004, corralDependenceEarthquakeRecurrence2006} in the description of earthquake waiting time statistics, and the SOC mechanism was introduced to account for the complexity of the earthquake \citep{Bak}. In this scenario, the waiting time statistics of earthquakes can be described by a universal scaling relation

\begin{equation}
\bar{\tau} \cdot P(t, M) \sim g(t/\bar{\tau}). \label{P_121102}
\end{equation}

The average waiting time $\bar{\tau}$ depends on magnitude $M$ and is defined as the sum of the intervals of earthquakes divided by the total numbers. The scaling function $g$ is the same for all magnitudes. A similar scaling argument is adopted here, in which $\bar{\tau}(I)$ provides an appropriate rescaling factor for the waiting time statistics, where $I$ represents the fluence of bursts, and will be excluded unless specifically required in further analyses.

The rescaling of all previous distributions by the average waiting time $\bar{\tau}$ at different fluence thresholds can be seen in Figure \ref{fig:frb121102}(b). All the data collapse onto a single curve, and the scaling function is

\begin{equation}
g(x) \sim (1 + \frac{x}{\beta})^{-\gamma}, \label{f_scale}
\end{equation}
where $\gamma = 2.22 \pm 0.23$ and $\beta = 0.60 \pm 0.08 $, respectively.

The above results show that a single scaling function can reasonably describe the waiting time distribution for different thresholds, with only small differences in each case. This function is also referred to as the generalized Pareto distribution \citep{hosking_parameter_1987}. Thus, the relation $g$ can be regarded as a unified scaling function, reflecting the self-similarity in intensity of the temporal occurrence of repeating FRBs. It is remarkable that, despite the large variability associated with the intensity of fluence, the temporal occurrence of repeating FRBs is governed by a simple, unique scaling law that is much more general and independent of any model of its temporal occurrence. Some deviations from the scaling function are observed for large $t$, as shown in Figure \ref{fig:frb121102}(b). For the typical minimum observing session, about 1 hour for one day, the bin range for the last data point is close to the interval of 1100 to 3200 seconds, while the time range for the second-to-last data point is close to 360 to 1100 seconds. One possible cause of this discrepancy may be attributed to the finite length of each observation session, which underestimates the frequency of long waiting times. In the future, more bursts of FRB data may improve the statistical behavior of waiting time in these intervals.

Unlike the longer peak, the position of the shorter peak remains unchanged with increasing fluence \citep{FAST1}, challenging the universal scaling law. The appearance of the longer peak is influenced by the activity level of the source, and to maintain consistency across all datasets, we focus on waiting times that fall within the scaling domain. This means considering only the waiting times that exceed the shorter peak interval, which is approximately 1 second for FRB 20220912A and FRB 20201124A (B), and approximately 10 seconds for FRB 121102 and FRB 20201124A (A). In the following, needing to uniformly express minimum time intervals, we consistently use 1 second as the minimum burst time interval. However, it must be clear that this is only the minimum value among the four sets of observed data that we focus on. Note that although the isotropic equivalent energy distribution is characterized by a bimodal structure \citep{FAST1}, this analysis clearly shows that all bursts collapse onto a single scaling function for waiting time, which is independent of energy thresholds.

We performed the same analysis on FRB 20220912A for comparison purposes. The waiting time distribution $P(t)$ exhibits similar behavior across various fluence thresholds, just as the results for FRB 121102 presented in Figure \ref{fig:frb121102}(a). Upon rescaling using Eq. (1), all thresholds in the waiting time distribution show consistent behavior with the same function $g(x)$ [Eq.(\ref{f_scale})], characterized by $\gamma = 2.63 \pm 0.15$ and $\beta = 0.82 \pm 0.15 $, respectively. Generally, different burst sources have very different burst environments and properties, but the statistical analysis findings of these two very extensive samples of repeating FRBs suggest the universally applicable scale invariant waiting time distribution. 

To further verify the universality of the waiting time distribution of FRBs, we performed a statistical analysis of the data from the two active burst periods of FRB 20201124A. It was found that there were explicit differences in the statistical parameters between the two time periods (separated by about 106 days), and this may be related to the burst mechanism. In the subsequent discussion section, we will compare this phenomenon with the different behaviors of solar flares during maximum and minimum periods, in the hope of gaining a deeper understanding of this difference.

\begin{figure}
 \centering
 \includegraphics[width=\columnwidth]{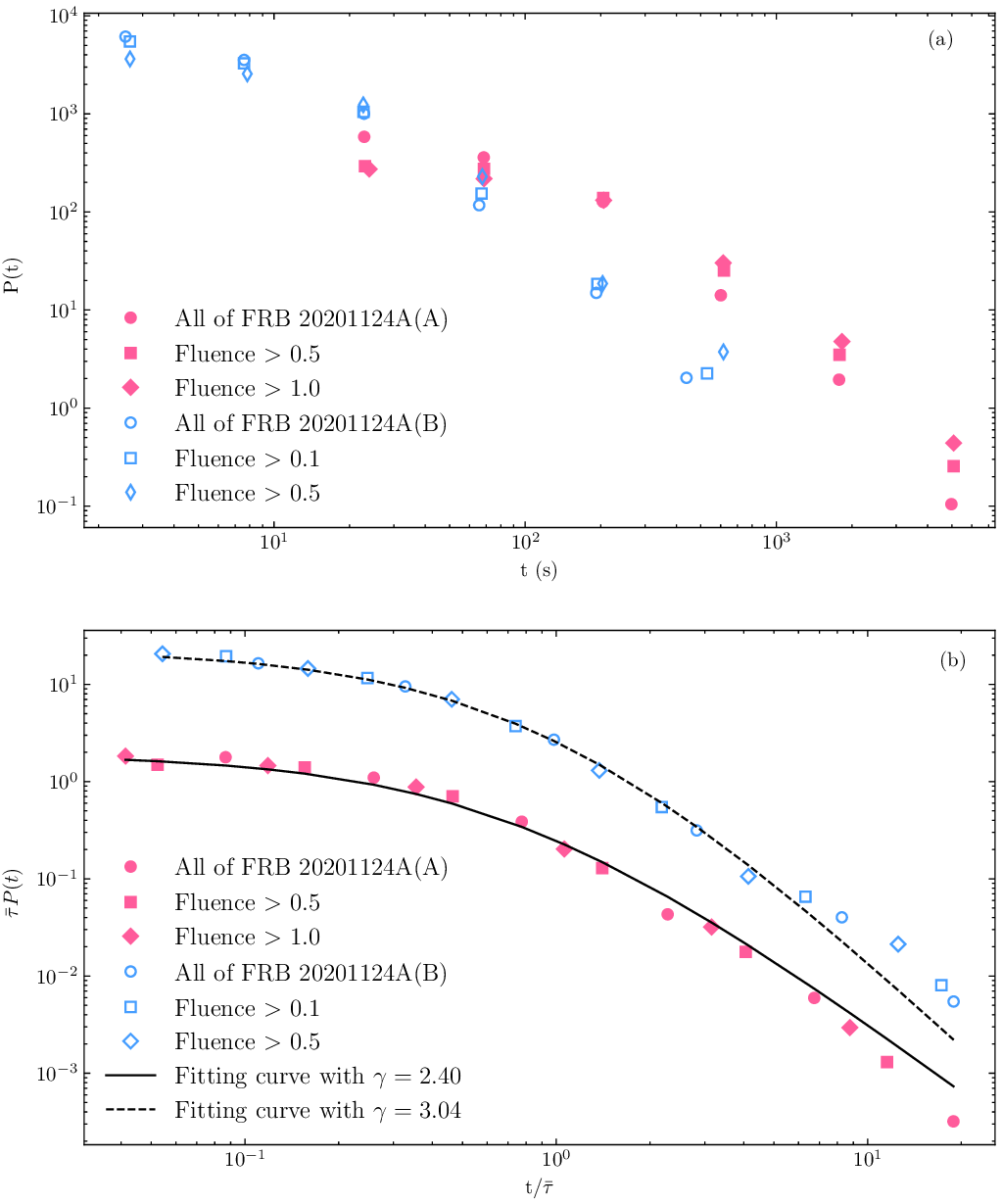}
 \caption{Comparison from the two active burst periods of FRB 20201124A. (a) Waiting time distributions $P(t)$ for different fluence thresholds (A: unit: Jy s, B: unit: Jy ms). (b) Waiting time distributions after rescaling by the average waiting time of bursts in the case of different fluence thresholds, respectively. For a clearer comparison, the data and fitting curve for FRB 20201124A(B) have been vertically shifted.}
 \label{fig:frb20201124A}
\end{figure}

As shown in Figure \ref{fig:frb20201124A}(a), before time rescaling, FRB 20201124A (B) exhibits a more active state on shorter time scales, while FRB 20201124A (A) shows relative activity on longer time scales. Similar to the behavior of FRB 121102 and FRB 20220912A, under different threshold values, both their waiting time distributions $P(t)$ display varying trends. After rescaling, the same unified function $g(x)$ [Eq.(\ref{f_scale})] applies to both FRB 20201124A (A) and FRB 20201124A (B). The parameters are $\gamma = 2.40 \pm 0.59$ and $\beta = 0.74 \pm 0.26$ for FRB 20201124A (A) and $\gamma = 3.04 \pm 0.56$ and $\beta = 0.95 \pm 0.24$ for FRB 20201124A (B). The presence of the same unified function in the description of the waiting time of FRB 20201124A both for two active periods is intriguing, suggesting that the function form of the waiting time distribution is universal. The difference in the parameter $\gamma$ may reveal more characteristics about the central engine and radiation mechanisms.

\begin{figure}
 \centering
 \includegraphics[width=\columnwidth]{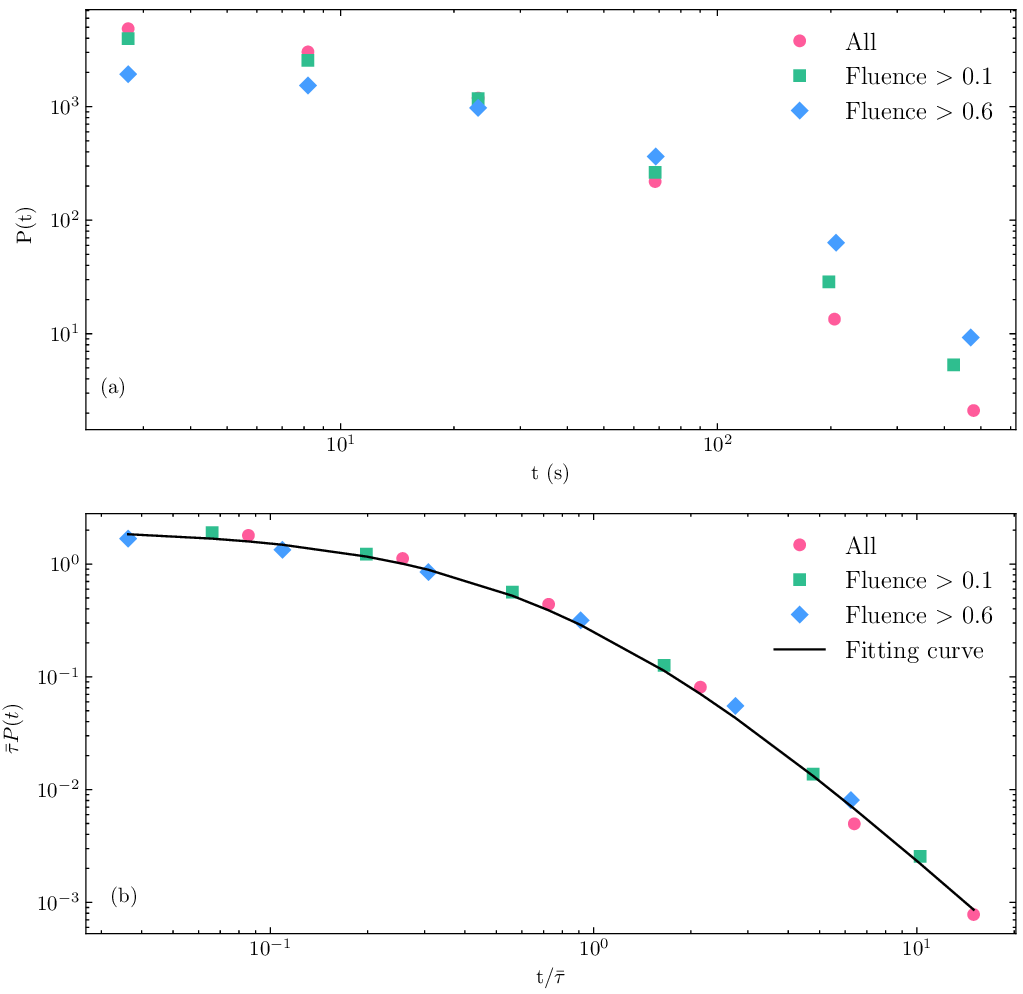}
 \caption{Waiting time distributions without and with rescaling of FRB 20220912A. (a) Waiting time distributions $P(t)$ for different fluence thresholds (Unit: Jy ms). (b) Waiting time distributions after rescaling by the average waiting time of bursts in the case of different fluence thresholds, respectively.}
 \label{fig:frb20220912A}
\end{figure}

The slow decay observed in the waiting time distribution [Eq.(\ref{f_scale})] indicates clustered bursts, unlike a Poisson process where bursts are evenly spread over time. This implies that bursts tend to occur close to each other over short periods. However, the lack of exponential decay might be due to the correlated system driving. Although the physical origin of this correlated drive is background dependent, our current analysis, which characterizes the temporal occurrence of bursts using a unified law, supports the existence of universal mechanisms in the burst generation process, governed solely by the burst occurrence rate $\lambda$, which is inverse to the average waiting time $\bar{\tau}$.

If the burst rate $\lambda (t)$ of repeating FRBs is assumed to vary slowly over time, this process can be employed to model time-dependent Poisson processes for the waiting time distribution of solar flares \citep{Wheatland, Aschwanden}, resulting in a description of the waiting time distribution as

\begin{equation}
P(t, I) = \frac{\int^{\infty}_0 f_I(\lambda)\lambda^2\exp{(-\lambda t)}d\lambda}{\overline{\lambda}_I}, \label{time_Possion}
\end{equation}

where the denominator $\overline{\lambda}_I = \int^{\infty}_0 \lambda f_I(\lambda)d\lambda$ is the average burst rate and $f_I(\lambda)d\lambda$ is the fraction of the time that the burst rate is in the range $(\lambda, \lambda + d\lambda)$ with intensity greater than the threshold $I$.

The analytic form of $f_I(\lambda)$ is constrained by the scaling function $g(x)$, when the analytic form of $f_I$ is taken as

\begin{equation}
f_I(\lambda) \sim \lambda^{-2} G_I(\lambda), \label{GI}
\end{equation}
and $G_I$ is the Gamma distribution:

\begin{equation}
G_I(\lambda) \sim (\frac{\beta\lambda}{R_I})^{\gamma-1}\exp(-\frac{\beta\lambda}{R_I}). \label{burst_rate}
\end{equation}
Here, $R_I$ is the inverse of the average waiting time, denoted as $\bar{\tau}$ in Eq.(\ref{P_121102}), and the unified scaling function Eq.(\ref{f_scale}) is restored after taking $f_I$ into Eq.(\ref{time_Possion}).

The characteristic of the non-Poisson waiting time distribution suggests that the central engine of bursts may involve a more complex dynamical mechanism \citep{wang_repeating_2023}, rather than a pure stochastic process. Recent studies indicate that a superposition of different dynamics can effectively describe many driven non-equilibrium systems with sufficient complexity across different time scales \citep{Manshour, Carbone, Beck, Beck1, Beck2}. Here, $\lambda$ in the Gamma distribution $G_I$ corresponds to the superposition of $n = 2\gamma$ independent Gaussian variables $X_k$, each with an average of 0 and a squared magnitude.

\begin{equation}
\lambda = \sum^n_{k=1}X^2_k. \label{Lambda_Gaussian}
\end{equation}

The Gamma distribution arises naturally for a variable with a limited number of degrees of freedom. For example, in the case of FRB 121102, this number is approximately $n \approx 4 $, and for FRB 20220912A, it is around $n \approx 5$. The precise value of the parameter $\gamma$ depends on the degrees of freedom of the underlying Gaussian distributions, which are determined by particular stochastic processes.

The unified scaling function $g(x)$ can also be rewritten as:

\begin{equation}
g(x) \sim (1 +\lambda_0(q-1)x)^{-\frac{1}{q-1}}, \label{q-exponent}
\end{equation}
after replacing $\gamma$ with $1/(q-1)$ and $\beta$ with $\gamma/\lambda_0$, and the derived function is termed the Tsallis q-exponential function. Here, $\lambda_0$ is defined as $\lambda_0 = \int_0^\infty \lambda G_I(\lambda)d\lambda$. The parameter q quantifies the deviation from a standard exponential distribution, where a value of $q>1$ suggests a long-tailed trend. Although our initial analysis of burst temporal complexity began with the assumption of exponential distribution, deviations from this standard were noted. Subsequent insightful transformations led us to adopt the q-exponential distribution.

\section{Discussion}

Solar flares display a comparable scaling characteristic in their waiting time distribution, which is independent of the intensity threshold throughout the solar cycle \citep{Baiesi}. During the solar maximum, the parameter $\gamma \simeq 2.83$. In contrast, during solar minimum, $\gamma \simeq 1.51$, which supports well the distribution observed in the laminar phase of marginal behavior in on-off intermittency, a model with an exponent of $\gamma = 3/2$ \citep{Heagy}. Different types of intermittency, as discussed by Schuster \citep{Schuster}, show varying exponents in the distribution of the laminar phase. For Type-II intermittency, the exponent is $\gamma = 2$, similar to the waiting time distribution parameter $\gamma \simeq 2.22$ observed for FRB 121102 and $\gamma \simeq 2.40$ for FRB 20201124A(A).

In comparison, the temporal characteristics exhibited by the same FRB 20201124A source during two different active outburst periods (approximately 106 days apart) show more similarities with the temporal characteristics of solar flares during solar maximum and minimum. Apart from their shared waiting time distribution functions and interpretations for different types of intermittency at smaller $\gamma$ values, the $\gamma$ value during solar maximum is closer to that of FRB 20201124A(B), being $2.83$ and $3.04$, respectively.

Further theoretical and observational studies have found that the power law index of the waiting time distribution for solar flares varies between 1.4 and 3.2 \citep{aschwanden_reconciliation_2010, aschwanden_correlation_2021, aschwanden_solar_2021}, and its value correlates with the solar activity cycle. The rate of flare variation can be spike-like or flatter. Theoretical studies have suggested that the size of the index characterizes the degree of nonlinearity in the temporal evolution of flare dynamics. Uniformity in time interval distributions among different FRB sources, irrespective of their fluence variations, highlights the importance of analyzing inter-event time statistics for comprehending burst dynamics. This finding is in contrast to the distribution of the burst energy. Such observations may point to the possibility of revealing crucial aspects of the burst mechanism through a statistical analysis of the event time distributions, as suggested by \citet{kumarIntereventTimeDistributions2020}.

Despite the fact that both earthquakes and solar flares are prominent examples in the field of SOC. However, the differences in their waiting time distributions are intriguing. First, while both conform to the universal scaling relation, their distribution functions exhibit different behaviors over longer time scales. Earthquakes' scaling function aligns closely with a generalized gamma distribution $g(x) \sim exp(-x^{\delta}/B)/x^{1-\gamma}$ \citep{corralLongTermClusteringScaling2004, corralDependenceEarthquakeRecurrence2006}, whereas solar flares are better represented by the generalized Pareto distribution [Eq.(\ref{f_scale})] \citep{hosking_parameter_1987}. The difference at shorter time scales (less than 1 second) is related to aftershocks, predominantly characterized by Omori-Utsu law statistics, which are evident in the generalized gamma distribution. The real divergence arises in the exponential component of the gamma function, which differs from the behavior of the Pareto distribution and acts as a power law at longer time scales. However, analyzing this interval involves changes in the event rate over longer periods and suffers from limited statistical data. Therefore, we must be careful to consider the comparison with earthquakes in this interval as a topic for future studies. Second, solar flares display varying parameters throughout different phases of the solar cycle, in contrast to earthquakes, which demonstrate nearly constant parameters that remain consistent across various spatial and temporal scales. Of course, there are potential reasons mentioned previously that long-term changes in the event rate could impact the variability of the parameters describing earthquakes. Both solar flares and repeating FRBs involve highly energetic explosions characterized by complex temporal patterns and rapid energy releases. A recent study, which focuses on threshold-independent analysis of waiting times, suggests that if the distribution of burst energy follows a power law in the background of FRBs or SGRs, it could potentially serve as an indicator of the dynamics of these phenomena \citep{katz_log-normal_2023}.

Although our analysis focuses on long-term behavior in the waiting time distribution of FRBs, we recognize that at shorter waiting timescales, FRB aftershock sequences may exhibit characteristics similar to earthquakes, as reported by \cite{totaniFastRadioBursts2023}. This reflects a potential similarity in the statistical laws that govern the short $t$ region of the waiting time distribution between FRBs and earthquakes, particularly given that earthquake aftershock sequences follow the Omori-Utsu law. Therefore, while our study acknowledges the similarity between FRBs and earthquakes, it emphasizes the need to explore various mechanisms to comprehend these occurrences on various temporal scales. This consideration is crucial when examining the high-energy explosions of FRBs or solar flares and their dynamics, as these processes may involve the convective transfer of magnetic energy from the interior to the exterior of stellar bodies. This mechanism is particularly significant on longer time scales.

The dynamics of magnetic energy within celestial bodies, from solar flares on the Sun to FRBs radiating out from neutron stars, potentially highlight a fundamental astrophysical process: the convective emergence of magnetic energy. This process, integral to the appearance of solar flares, involves the transport and intricate reconfiguration of magnetic field lines at the surface, leading to magnetic reconnection that releases significant energy \citep{parkerHydromagneticDynamoModels1955, charbonneauDynamoModelsSolar2020}. Herein, we extend this understanding to the neutron star context, proposing a unified framework that possibly captures the essence of both phenomena and offers insights into the complex dynamics of magnetar surrounding FRB occurrences.

Our investigation reveals that the scaling properties of repeating FRBs over longer time intervals (i.e. greater than approximately 1 second) resemble those seen in solar flares, suggesting a similarity in the underlying mechanism. In contrast, FRBs with shorter intervals (that is, less than about 1 second) display distinctly different scaling properties, associating more closely with earthquake dynamics \cite{totaniFastRadioBursts2023}. This divergence is centered on a threshold time scale of approximately 1 second, which is notably comparable to the dynamic time scale of neutron star crusts. The convective processes within a neutron star can alter the internal magnetic field configuration, affecting superfluid vortices' pinning energies and, consequently, the time scales for crust fractures. Such fractures, which are believed to result from stress release due to differential rotation between the superfluid core and the solid crust, occur on shorter time scales than the observed rise times of glitches \citep{sideryEffectQuantizedMagnetic2009}. The dynamics time scale, exemplified by the typical Vela pulsar glitch occurring in about 12.6 seconds \citep{ashtonRotationalEvolutionVela2019}, provides an estimate for the neutron star's crustal reconfiguration following significant fractures.

On the basis of these observations, we propose a scenario wherein the convective magnetic process within a neutron star leads to crust fractures, releasing energy, and producing FRB bursts. This mechanism's similarity to solar flare processes is reflected in the comparable scaling properties. Following the fracture, the crust may undergo aftershock-like activities that last a few seconds, some powerful enough to generate additional FRB bursts. These findings suggest that crust fractures related to FRBs are likely smaller than those causing glitches, as no temporal coincidence with time glitches has been observed, and that the predominant mechanism draws a convincing similar to solar flare dynamics. According to Bak \citep{Bak, Bak_wt}, the observed scaling law in burst events, especially in the temporal clustering of recurrence bursts, indicates a complex spatio-temporal pattern, possibly indicating an intermittent energy release from a self-organized system, which aligns with the theoretical framework of critical phenomena. 

The observed scaling parallels between repeating FRBs and solar flares illuminate a probable shared mechanism rooted in the convective release of magnetic energy within neutron stars. This insight challenges the conventional emphasis on magnetosphere dynamics alone. Instead, our findings support a general view that involves both magnetic and convective activities within neutron stars as key drivers. The dynamic interplay observed around magnetar, particularly in relation to FRB occurrences, incorporates seismic-like aftershocks at shorter intervals with convective magnetic energy transfers over longer durations. This comprehensive framework, aligned with broader magnetar theory and highlighted by the insights from \cite{totaniFastRadioBursts2023}, not only deepens our understanding of FRBs and solar flares but also emphasizes the complex multiscale nature of magnetar behavior. This approach encourages further exploration into the convective and magnetic processes within neutron stars, promising to reveal the intricate dynamics underlying these mysterious phenomena.

\section{Summary}

Although we have extensively analyzed and compared the similarities in the burst timing characteristics of repeating FRBs, solar flares, and earthquakes, we cannot exclude the possibility of other burst mechanisms. Similar statistical results have been observed for certain observable quantities (such as velocity or magnetic fluctuations) and waiting time distributions in fully developed turbulence \citep{Beck2, Beck, Manshour}. Intermittency in turbulence occurs in fluids where chaotic dynamics and power law statistics coexist. This contrasts with SOC processes, where the phenomenon of intermittency is explained by conceptually different mechanisms.

In summary, recent observations by FAST of repeating FRBs, particularly FRB 121102, FRB 20220912A and FRB 20201124A, have revealed high burst rates and unique energy distribution and temporal characteristics. In this study, we conducted a comprehensive analysis of the temporal occurrence of these repeating FRBs, revealing a unified scaling law for waiting times between bursts longer than approximately 1 second. This scaling law exhibits similarities to that of solar flares at this longer timescale, leaving the similarity to earthquakes in this range as a subject for future research. We propose a dual analogy for the FRB scenario, with similarities to earthquake dynamics at subsecond scales and solar flare dynamics beyond the one-second threshold. This analogy offers new insights of the time clustering of repeating FRBs into the internal processes of neutron stars and suggests a complex, multifaceted dynamics that governs repeating FRBs. Evidence of scale invariance across different fluence thresholds for these FRB sources also suggests a complex underlying dynamic mechanism influenced by SOC. Future research aimed at probing deeper into the connections among the temporal patterns of repeating FRBs, solar flares, and earthquakes will not only create a consistent link across different astrophysical events, but also enhance our understanding of the underlying dynamics that drive the occurrences of repeating FRBs.

\section*{Acknowledgements}

We appreciate the anonymous referrer for insightful suggestions. We thank Mingyu Ge for useful discussions. 
We thank the support from the National Key R\&D Program of China (Grant No. 2021YFA0718500), 
the National Natural Science Foundation of China 
(Grant No. 12273042 
) and
the Strategic Priority Research Program of the Chinese Academy of Sciences (Grant No. XDB0550300, 
XDA30050000,   
XDA15360000). 
We acknowledge the computing resources provided by the National High Energy Physics Data Center and the High Energy Physics Data Center, Chinese Academy of Science. 

\section*{Data Availability}

All data analyzed in this paper are included in references \citep{FAST1, FAST2, zhangFASTObservationsFRB2023, zhouFASTObservationsExtremely2022} separately. They are also openly available in the Science Data Bank at https://doi.org/10.11922/sciencedb.01092 \citep{FAST1}, https://cstr.cn/31253.11.sciencedb.08058 \citep{zhangFASTObservationsFRB2023}, https://cstr.cn/31253.11.sciencedb.06762 \citep{zhouFASTObservationsExtremely2022}, and on the PSRPKU website, https://psr.pku.edu.cn/index.php/publications/frb20201124a/ \citep{FAST2}.



\bibliographystyle{mnras}
\input{article.bbl} 





\bsp	
\label{lastpage}
\end{document}